\documentclass{aa}
\usepackage{graphicx}
\usepackage{natbib}
\usepackage{txfonts}
\bibpunct{(}{)}{;}{a}{}{,} % to follow the A&A style     
\begin{document}
\title{XMM-Newton spectral and timing observations of the millisecond
  pulsar \object{PSR J0218+4232} }

   \subtitle{}

   \author{N.A. Webb
          \and
          J.-F. Olive
          \and
          D. Barret
          }

   \offprints{N.A. Webb, \email{Natalie.Webb@cesr.fr}}

   \institute{Centre d'Etude Spatiale des Rayonnements, CNRS/UPS, 9 Avenue du Colonel Roche, 31028 Toulouse Cedex 04, France }

   \date{Received <date> / Accepted <date>}

   \abstract{We present XMM-Newton MOS imaging and PN timing data of
   the millisecond pulsar \object{PSR J0218+4232}.  We confirm the
   previously detected pulsations of \object{PSR J0218+4232} and we
   show that the folded lightcurve is dependent on energy.  We present
   the broad band (0.2-10.0 keV) spectrum of this millisecond pulsar,
   as well as the spectra of: the pulsed emission, the individual
   pulses, the interpulse and the non-pulsed region and we compare our
   results with previous data from Rosat, BeppoSAX and Chandra.  We
   discuss the results of the spectral fitting in the context of
   recent pulsar models.

     \keywords{X-rays: stars -- pulsars: individual: PSR J0218+4232 --
       Radiation mechanisms: non-thermal -- Radiation mechanisms:
       thermal} }

   \maketitle
%
%________________________________________________________________

\section{Introduction}

Pulsars show two different types of X-ray pulsations: soft pulsed
emission with blackbody spectra where the emission is from the
neutron star surface, such as cooling emission from the whole surface
as a relic of the neutron star's birth and/or the emission from a
heated region, most likely the polar cap; or pulsed emission with
narrow pulses and hard power law spectra, where it is thought that the
emission arises from the magnetosphere \citep[see e.g.][for further
discussion]{sait97,kuip98,taka01}.  Therefore detecting not only a
pulsation, but also determining the form of the pulse profile, could
help to identify the underlying emission mechanism and detection of
the polar cap thermal radiation would allow us to discriminate between
various models of pulsars and study the properties of neutron star
surface layers \citep{zavl02,pavl97,zavl98}.  Many millisecond pulsars
(MSPs) such as \object{PSR B1821-24} \citep[e.g.][]{sait97} and
\object{PSR B1937+21} \citep[e.g.][]{taka01}, like many ordinary
pulsars, show X-ray pulsations and have been observed to emit X-ray
spectra that are well fitted by a hard power law.  In contrast, it has
been proposed that the millisecond pulsar \object{PSR J0437-4715}
emits thermal radiation from hot polar caps \citep{zavl98,zavl02}.

The radio source \object{J0218+4232} has been known for many years
\citep{dwar90,hale93} although it was first confirmed to be a
millisecond pulsar by \cite{nava95}, using radio observations made
with the Lovell telescope in 1993.  A pulse period of 2.3 ms was
determined.  \cite{nava95} also showed that this luminous pulsar
\citep[L$_{400} > 2700$ mJy kpc$^2$][where L$_{400}$ is the luminosity
at 400 MHz]{nava95} is in orbit with a low mass white dwarf
(~0.2M$_\odot$), with an orbital period of about two days.  From the
dispersion measure a lower limit on the distance of 5.7 kpc was
derived.

\cite{verb96} showed evidence for \object{PSR J0218+4232} at high
energies, both in X-rays using ROSAT and $\gamma$-rays, using EGRET,
the High-Energy Gamma-Ray Telescope aboard the Compton Gamma-Ray
Observatory (CGRO).  They also presented some indication of pulsation
in both energy domains.  \cite{kuip00} provided stronger evidence for
the detection of pulsed $\gamma$-ray emission from \object{PSR
J0218+4232}, also using EGRET data.  \cite{kuip98} presented
further evidence for X-ray pulsation at the radio pulse period using
98 ks of ROSAT data.  They also likened the sharp pulses to those of
\object{PSR B1821-24} and the
\object{Crab pulsar}, which indicates a magnetospheric origin of the
pulsed X-ray emission.  They also estimated that almost two thirds of
the soft X-ray (0.1-2.4 keV) emission is non-pulsed, where
\cite{nava95} also stated that a large fraction of the radio emission
is not pulsed.  This lead them to the conclusion that \object{PSR
J0218+4232} is an aligned rotator.  This was further supported by
\cite{stai99}, using polarimetric radio observations.

BeppoSAX observations of \object{PSR J0218+4232} \citep{mine00}
provided detailed information on the pulsar's temporal and spectral
emission properties (1.6-10 keV).  They showed that pulse 1
\citep[lying at $\phi$=0.8, in][]{mine00} is much stronger than pulse
2 (lying at $\phi$=0.3) in the 1.6-4 keV energy band, but that pulse 2
is much stronger than pulse 1 in the 4-10 keV energy band.  Also at
the higher energy no non-pulsed component was apparent above the
background level.  They determined a very hard power law fit, photon
index = 0.61$\pm$0.32, to the pulsed part of the spectrum, with a
fixed column density of 5$\times 10^{20}$ cm$^{-2}$
\citep[see][]{verb96}, which gave a flux of 4.1$\times 10^{-13}$ ergs\
cm$^{-2}$ s$^{-1}$ (2-10 keV).  The measured spectral indices of
pulse 1 and pulse 2 were 0.84$\pm$0.35 and 0.42$\pm$0.36.  The total
spectrum was well fitted by a power law, with a photon index of
0.94$\pm$0.22 and they found a flux of 4.3$\times 10^{-13}$ ergs
cm$^{-2}$ s$^{-1}$ (2-10 keV).  \cite{kuip02} showed that the
non-pulsed spectrum of \object{PSR J0218+4232} has a softer spectrum
than the pulsed emission, using 73 ks of Chandra data.

Recently, \cite{kuip02,kuip02b} presented multiwavelength (radio, X-
and $\gamma$-ray) pulse profiles of \object{PSR J0218+4232}.  They
found that the two X- and $\gamma$-ray pulses are aligned and that the
X-ray pulses are also aligned with two of the three radio pulses.
This is a very interesting result as multiwavelength emission can help
to discriminate between the different theoretical models proposed to
describe the nature of millisecond pulsars, such as the polar cap
models \citep[see][and references therein]{hard02} and the outer gap
models \cite[see][and references therein]{chen00}, and thus trace the
origin of the emission.  For example, in polar cap scenarios,
$\gamma$-ray pulses should be accompanied by X-ray pulses at the same
phase and with similar shapes \citep{ruda99}.

XMM-Newton combines the spectral capabilities of ROSAT (0.1-2.5 keV)
and the BeppoSAX MECS (2-10 keV), which, when coupled with the
XMM-Newton's sensitivity, gives us unparalleled spectral information
from low to high energies.  We have thus used XMM-Newton to observe
the millisecond pulsar \object{PSR J0218+4232} to show the evolution
of the pulsed profile from low energies (0.4-1.6 keV) to high energies
(up to 12 keV).  We present spectra of the pulsar as a whole and both
the pulsed and non-pulsed components and we draw some conclusions
about the nature of the emission from \object{PSR J0218+4232}.

\section{Observations and data reduction}

\object{PSR J0218+4232} was observed by XMM-Newton on 2002 February
11-12.  The observations spanned 37.2 ks (MOS cameras) and 36 ks (PN
camera), but a soft proton flare affected 17 ks of the MOS exposure
and 16.5 ks of the PN exposure.  The MOS data were reduced using
Version 5.3.3 of the {\it XMM-Newton} SAS (Science Analysis Software).
However, for the PN data we took advantage of the development track
version of the SAS.  Improvements have been made to the OAL (ODF
(Observation Data File) access layer) task (version 3.106) to correct
for spurious and wrong time values, premature increments, random jumps
and blocks of frames stemming from different quadrants in the timing
data in the PN auxiliary file, as well as correcting properly for the
onboard delays \citep{kirs03}.  Using this version, our timing
solution improved (see Sect.~\ref{sec:pntiming}).

We employed the MOS cameras in the full frame mode, using a thin
filter \citep[see][]{turn01}.  The MOS data were reduced using
`emchain' with `embadpixfind' to detect the bad pixels.  The event
lists were filtered, so that 0-12 of the predefined patterns (single,
double, triple, and quadruple pixel events) were retained and the high
background periods were identified by defining a count rate threshold
above the low background rate and the periods of higher background
counts were then flagged in the event list.  We also filtered in
energy. We used the energy range 0.2-10.0 keV, as recommended in the
document `EPIC Status of Calibration and Data Analysis'
\citep{kirs02}.  The event lists from the two MOS cameras were merged,
to increase the signal-to-noise.

The PN camera was also used with a thin filter, but in timing mode
which has a timing resolution of 30$\mu$s \citep{stru01}.  The PN data
were reduced using the `epchain' of the SAS.  Again the event lists
were filtered, so that 0-4 of the predefined patterns (single and
double events) were retained, as these have the best energy
calibration and we filtered in energy.  The document `EPIC Status
of Calibration and Data Analysis' \citep{kirs02} recommends use of PN
timing data above 0.5 keV, to avoid increased noise.  We used, in
general, the data between 0.6-12.0 keV as this had the best
signal-to-noise, although we have also tried to exploit the data in
the 0.4-0.6 keV band (see Sect.~\ref{sec:pntiming}). The on-board
event times expressed in the local satellite frame were subsequently
converted to Barycentric Dynamical Time, using the SAS task `barycen' and
the coordinates derived from 6.5 years of radio timing measurements,
as given in \cite{kuip02}.

\section{Timing analysis}
\label{sec:pntiming}

In the PN timing mode, all the two-dimensional spatial information is
collapsed into a single dimension (x-direction). We extracted the
pulsar data using a rectangular region centered on the pulsar.

\begin{figure}
   \includegraphics[width=8cm]{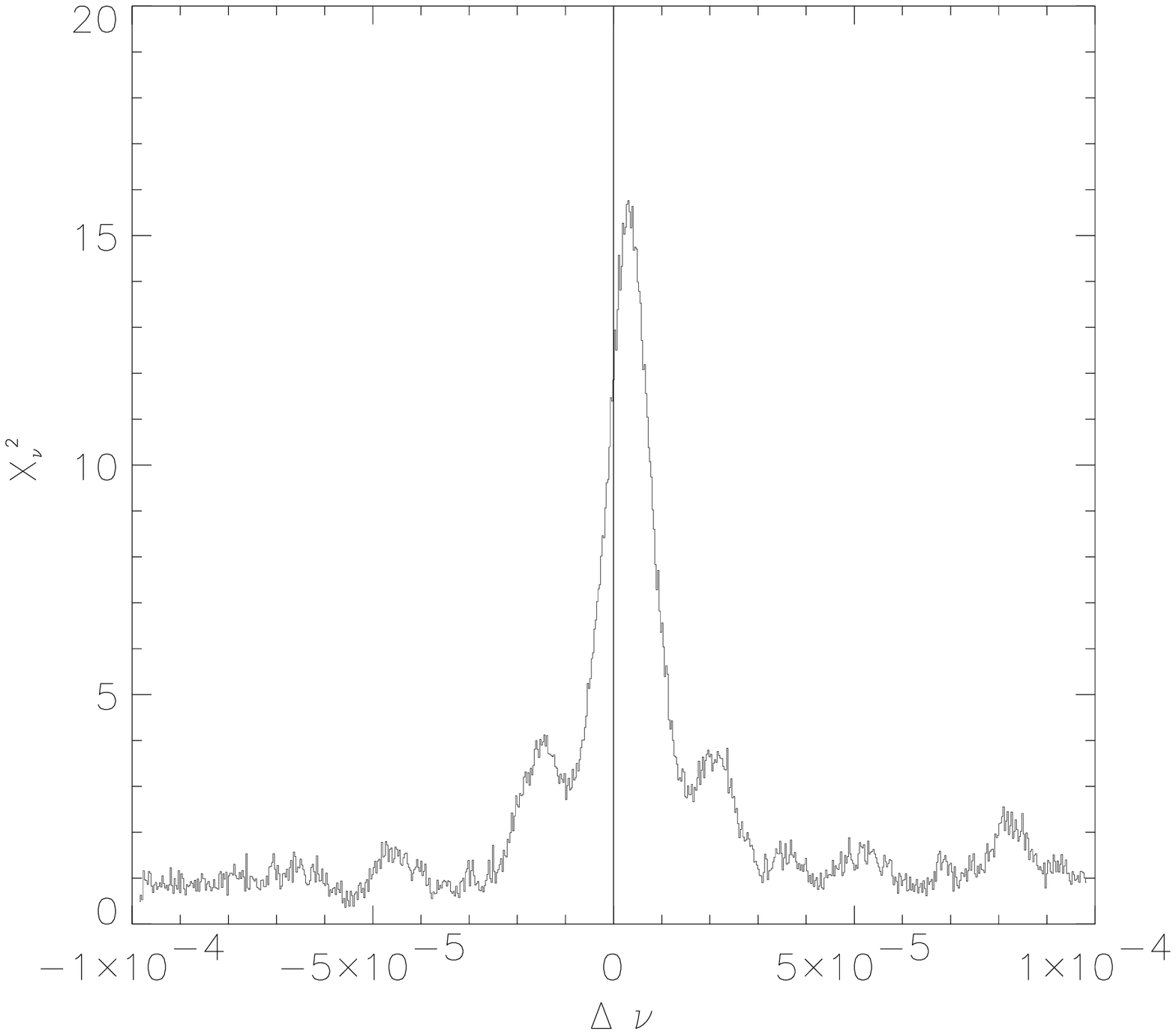}
   \caption{$\chi^{\scriptscriptstyle 2}_{\scriptscriptstyle \nu}$
   versus change in frequency from the expected pulsation frequency
   (shown as the solid vertical line at $\Delta \nu$ = 0.0),
   calculated with the ephemeris of \cite{kuip02}.  See
   Sect.~\ref{sec:pntiming} for details.}  \label{fig:chisquare}
\end{figure}

\begin{figure}
   \includegraphics[width=8cm]{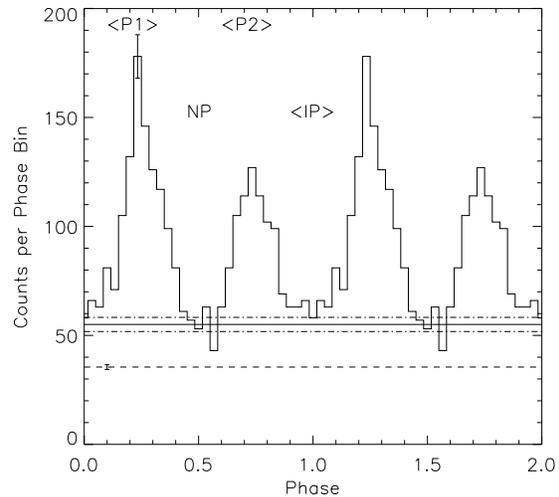}
   \caption{Lightcurve folded on the ephemeris of \cite{kuip02} and
   binned into 30 bins, each of 77$\mu$s.  Two cycles are shown for
   clarity.  Pulses 1 (P1) and 2 (P2) are indicated.  Between these
   two pulses is the section defined as the non-pulsed region (NP).
   Between pulses 2 and 1 is the section defined as the interpulse
   (IP).  A typical $\pm$1$\sigma$ error bar is shown.  The dashed
   line shows the background level, where the error bar represents the
   $\pm$1$\sigma$ error.  The solid line shows the DC level and the
   dashed-dotted lines represent the $\pm$1$\sigma$ error.  }
   \label{fig:folded}
\end{figure}

We corrected the data for the orbital motion of the pulsar
and the data were folded on the radio ephemeris given in
\cite{kuip02}.  We verified that the corrections made to the
timing data with the development version of the SAS are indeed
correct.  We tested frequencies at and around the expected frequency.
We used a $\chi^2$ test to determine the frequency at which the pulse
profile was the strongest. We found the largest peak in the
$\chi^{\scriptscriptstyle 2}_{\scriptscriptstyle
\nu}$ versus change in frequency from the expected frequency at
$\Delta \nu$ = 2 $\times 10^{-6} {\rm s}^{-1}$, see
Fig.~\ref{fig:chisquare}.  This is well inside the resolution of this
dataset ($\sim$1/T$_{obs}$), which is 3 $\times 10^{-5} {\rm s}^{-1}$,
thus we can conclude that the data reduction and analysis made to the
dataset are reliable.  The folded lightcurve (0.6-12 keV), data folded
on the radio frequency, counts versus phase, is shown in
Fig.~\ref{fig:folded}.  This lightcurve is very similar to the
lightcurve presented by
\cite{kuip02}, where both pulses can be found at approximately the
same phase \citep[within the timing uncertainties of Chandra and
XMM-Newton][]{kuip02,tenn01,kirs03}.  Fitting the two pulses with two
Lorentzians and a background \citep[in a similar way to][]{kuip02}, we
find a pulse separation of $\phi$=0.491$\pm$0.022 (90\% confidence
limit), consistent with previous results e.g. \cite{kuip98},
\cite{mine00} and \cite{kuip02}.  The FWHM of the pulses are
$\delta\phi_1$=0.112$\pm$0.038 for the pulse centred at phase
$\phi_1$=0.242$\pm$0.008 and $\delta\phi_2$=0.121$\pm$0.056 for the
pulse centred at $\phi_2$=0.733$\pm$0.014.  These values are also
consistent with the values presented in \cite{kuip02}.  Fitting the
pulses with two Gaussians or fitting the pulses separately gives a
similar result.

To calculate the DC (non-pulsed) level of our lightcurve we used the
bootstrap method outlined by \cite{swan96}.  The DC level is shown in
Fig.~\ref{fig:folded} by the solid line and the two dashed-dotted
lines show the $\pm1\sigma$ errors.  The background level (also shown
in Fig.~\ref{fig:folded}, dashed line) was determined by extracting
the data from a region in the same observation as the pulsar, centered
12 pixels ($\sim$50\arcsec) in the x-direction from the pulsar, where
there were no other X-ray sources close to this position or in the
full length of the y-direction.  We verified that there were indeed no
sources in this region, using the MOS imaging data (see
Section~\ref{sec:mos}).  The extraction region chosen for the
background was the same shape and size as the extraction region for
the pulsar.  We see evidence for both a DC interval between $\phi$ =
0.45-0.58 (see Fig.~\ref{fig:folded}) and an interpulse between $\phi$
= 0.88-1.11, in similar places to those of
\cite{kuip02}, where the interpulse region has 10.5$\pm$3.3 counts/bin
more than the non-pulsed region.  Thus we conclude that our X-ray
pulse profile can be compared directly to the lightcurve of
\cite{kuip02}.  We therefore refer to the pulse situated at
$\phi_1$=0.24 as pulse 1 (phase = 0.11-0.45) and $\phi_2$=0.73 as
pulse 2 (phase = 0.58-0.88) in the same way as \cite{kuip02}.
Subtracting the background from the total count distribution measured
in the strip centred on the pulsar, we found a pulsed percentage of
69$\pm$6\% in the 0.6-12.0 keV band, similar to the value determined
by \cite{kuip02}, who found 64$\pm$6\% with Chandra in the 0.08-10.0
keV band.  We find that pulse 1 contributes 65$\pm$7\% of the pulsed
percentage and pulse 2 contributes 32$\pm$8\%, where the ratio between
the pulse 1 and the pulse 2 counts is 1.60$\pm$0.09 in the
0.6-12.0 keV band.

%hardest band (4.0-10.0 keV).  This could be due to the smaller
%effective area of the BeppoSAX MECS at higher energies.

%\begin{figure}
%   \includegraphics[angle=-90,width=8cm]{0218lorentzians.ps}
%      \caption{The folded and binned lightcurve fitted with two 
%       Lorentzians and a background.   }
%   \label{fig:2peaksfit}
%\end{figure}                

\begin{figure}
\begin{flushleft}
    \includegraphics[width=10cm]{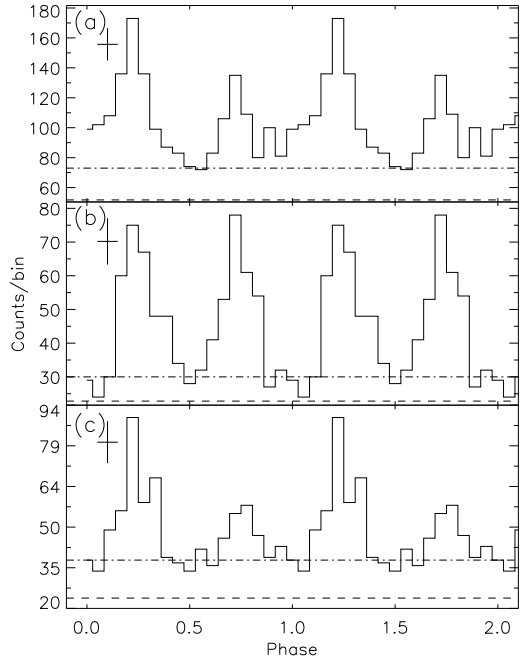}
      \caption{Lightcurves folded on the pulse period and binned into 
        18 bins.  (a) 0.4-1.6 keV (b) 1.6-4.0 keV (c) 4.0-10.0 keV.
        Two cycles are shown for clarity.  The dashed
        line indicates the background level, the dashed-dotted line
        indicates the DC level and an error bar represents the typical
        $\pm$1$\sigma$ error.  }
\end{flushleft}
    \label{fig:psrdiffenergy}
\end{figure}

Given the large number of counts in our observations, we were also
able to analyse the variations in the lightcurves as a function of
energy.  We chose the energy bands: 0.4-1.6 keV; 1.6-4.0 keV; and
4.0-10.0 keV, so that we could compare our results to those of
\cite{mine00}, see Fig.~\ref{fig:psrdiffenergy}.  The data were folded
as before and binned into 18 bins, where two cycles are presented for
clarity.  

The first thing that we notice from looking at
Fig.~\ref{fig:psrdiffenergy} is that the lightcurves are remarkably
different in the 3 energy bands, indicating that the spectra of
the pulses are indeed quite dissimilar.  Pulse 1 is always the
strongest of the two pulses, but the ratio between the number of
counts in the two pulses in the different energy bands varies quite
dramatically.  This can be seen quantitatively in
Table~\ref{tab:pulsedfracs}.  The softest and intermediate bands
are consistent with the ratios found by
\cite{mine00}.  However \cite{mine00} found that pulse 2 was stronger
than pulse 1 in the hardest band (0.4-10.0 keV), but they use a pulse
width of $\phi$=0.37 for pulse 2, whereas we use a pulse width of
$\phi$=0.30, which could account for the discrepancy.  We also see
evidence for the interpulse in the softest energy band.

As the ROSAT and BeppoSAX phasograms \citep{kuip98,mine00} are
not in absolute phase, it is important to determine which of the
pulses correspond to our pulse 1 and 2.  \cite{kuip00} present the tentative
alignment after cross-correlation of the two X-ray lightcurves (ROSAT HRI, 0.1-2.5 keV and BeppoSAX MECS 1.6-10.0 keV), where pulse 1 (the largest pulse in
both the ROSAT and the BeppoSAX lightcurves) appears at approximately
$\phi$=0.2, as in our lightcurve and that of \cite{kuip02}, which is
in absolute phase.  Thus we can compare pulses 1 and 2 in
\cite{mine00} with pulses 1 and 2 in this work.

Comparing our lightcurves with those presented in \cite{mine00}, we
find a remarkable visual difference.  In the intermediate energy band
(1.6-4.0 keV), they find only one clear peak - pulse 1.  However, we
find that we have two pulses of similar strength.  In the hardest
band, we see that pulse 1 is strongest, in contrary to the BeppoSAX
observations \citep{mine00}.  The discrepancies have yet to be
explained, but they may be related, to some extent, to differences in
the energy response of the different instruments.  In the softest band
(0.4-1.6 keV), we see that pulse 1 is strongest, similar to ROSAT
(0.1-2.4 keV), \citep{kuip98,mine00}.  \cite{kuip02b} also find
evidence that pulse 2 is stronger at higher energies than pulse 1,
using RXTE data in the range 2-16 keV, as well as some evidence that
pulse 2 is stronger than pulse 1 in the 2-8 keV range.  Examining the
8-12 keV range, we find that pulse 2 is 1.5 times stronger than pulse
1.  From this we have seen that the two pulses are spectrally quite
dissimilar.  With this in mind we investigated the pulse' spectra, as
well as other regions of the lightcurve (see Sect.~\ref{sec:specan}).

\begin{table}
\caption{Table showing the pulsed percentage and the ratio of pulse 1 to pulse 2 for the \object{PSR J0218+4232} in different energy bands. }
\label{tab:pulsedfracs}
\begin{center}
\begin{tabular}{ccc}
\hline
Energy Band (keV) & Pulsed \% & $\frac{pulse1}{pulse2}$ \\ 
\hline
0.6-12.0 & 69$\pm$6 & 1.60$\pm$0.1 \\
0.4-1.6 & 59$\pm$7 & 2.1$\pm$0.1 \\
1.6-4.0 & 68$\pm$11 & 1.1$\pm$0.1 \\
4.0-10.0 & 43$\pm$12 & 2.5$\pm$0.2 \\
\hline

\end{tabular}

\end{center}
\end{table}

\section{The MOS cameras' field of view}
\label{sec:mos}

The sources in the MOS field of view were first detected using the SAS
EPIC source detection task `eboxdetect', which employs a `local'
source detection algorithm.  A box of 5 $\times$ 5 pixels was used to
detect point sources and then the same box was used on the background.
To detect extended sources, two iterations were made with a box of 10
$\times$ 10 pixels and then 20 $\times$ 20 pixels, however only point
sources were detected.  A detection likelihood could then be
calculated.
 
\begin{figure}
   \centering
   \includegraphics[angle=0,width=8cm]{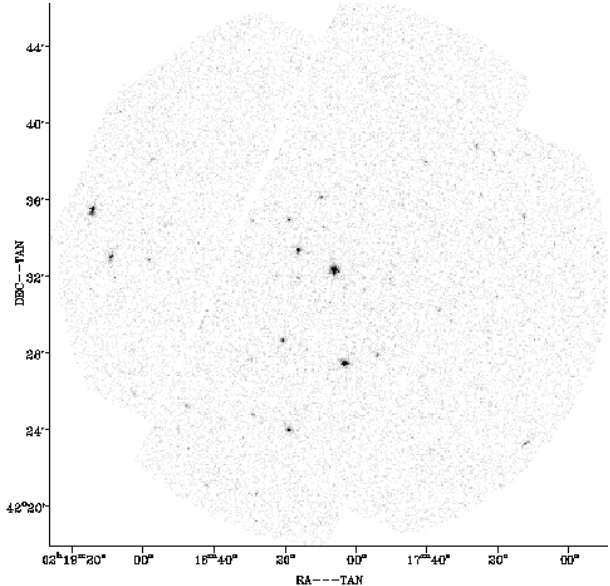}
      \caption{MOS (20.5 ks) image of the pulsar 
        field of view.  The pulsar can be seen as the
        brightest source in the centre of the field of view.}
   \label{fig:fov}
\end{figure}

42 sources have been detected using the SAS task `emldetect', using a
maximum detection likelihood of 18 (approximately 4$\sigma$ likelihood
of detection) and ignoring those sources not found on both cameras
(unless they lay outside the field of view or were in a chip gap).  We
calculate an unabsorbed flux limit of 1.0 $\times 10^{-14} {\rm ergs\
cm}^{-2} {\rm s}^{-1}$ (0.2-10.0 keV), using PIMMS (Mission Count Rate
Simulator) Version 3.2, with a power law, photon index of 2
\citep[as][]{hasi01}.  \cite{kuip02} found 51 sources in a comparable
field of view.  We detected 14 sources within the central 5\arcmin\
radius (excluding the pulsar), where \cite{mine00} found 7, using the
ROSAT data of \cite{kuip98}.  The whole field of view can be seen in
Fig.~\ref{fig:fov}.

\section{Spectral analysis}                     
\label{sec:specan}

The total energy spectrum of the MSP \object{PSR J0218+4232} can be
extracted from the MOS imaging data.  However, as the lightcurves have
been seen to vary dramatically from energy band to energy band (see
Sect.~\ref{sec:pntiming}), we have also used the PN timing data from
which the phase resolved spectral information can be extracted, to
isolate the spectra from different regions in the lightcurve.  We have
fitted these spectra with a variety of models, in an attempt to find
which is the most appropriate.

\subsection{MOS spectral analysis}

We extracted the MOS spectra using an extraction radius of
$\sim$1\arcmin\ and rebinned the data into 15 eV bins.  We used a
similar neighbouring surface, free from X-ray sources to extract a
background file.  We used the SAS tasks `rmfgen' and `arfgen' to
generate a `redistribution matrix file' and an `ancillary response
file'.  We then used Xspec (Version 11.1.0) to fit the spectra. We
binned up the data to contain at least 50 counts/bin.  Fitting the MOS
data together, we find that a good fit for the spectrum (between
0.2-10.0 keV) is a single absorbed power law, photon index of
1.10$\pm$0.06, $\chi^{\scriptscriptstyle 2}_{\scriptscriptstyle
\nu}$=1.00 (29 degrees of freedom (dof)), when the N$_{\hbox{\rm H}}$
was frozen at 5$\times 10^{20} {\rm cm}^{-2}$, see \cite{verb96}.
Allowing the N$_{\hbox{\rm H}}$ to vary, gives N$_{\hbox{\rm
H}}$=(7.6$\pm4.3)\times 10^{20} {\rm cm}^{-2}$ and a photon index of
1.19$\pm$0.12, $\chi^{\scriptscriptstyle 2}_{\scriptscriptstyle
\nu}$=1.04 (28 dof).  A blackbody plus a power law also provides a
good fit to the data (see Table~\ref{tab:pnspectralfits}, where the
unabsorbed flux (${\rm ergs\ cm}^{-2} {\rm s}^{-1}$) in the 0.6-10.0
keV range can also be found).  A single blackbody gives a very poor
fit to the data, kT=0.85$\pm$0.04 keV, $\chi^{\scriptscriptstyle
2}_{\scriptscriptstyle \nu}$=3.79 (29 dof), when the N$_{\hbox{\rm
H}}$ was frozen at 5$\times 10^{20} {\rm cm}^{-2}$.

\subsection{PN spectral analysis}

\subsubsection{The total pulsar spectrum}

\begin{figure}
   \centering \includegraphics[angle=0,width=9cm]{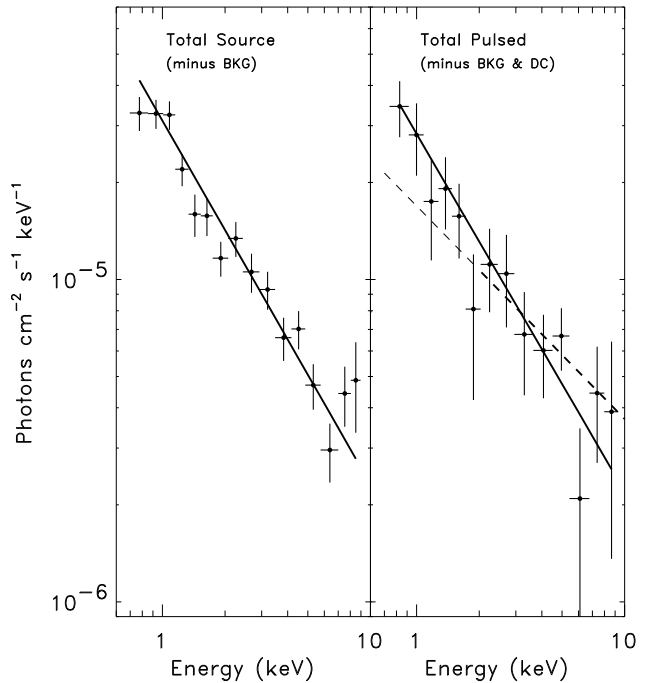}
   \caption{The left hand plot shows the power law fit (solid line) to
   the unfolded total (pulsed + non-pulsed) pulsar spectrum using EPIC
   PN data.  On the right hand side, the unfolded spectrum from the
   pulsed region only and the best fit to these data are shown.  The
   thick dashed line is the best fit to the data above 2 keV, see
   Section~\ref{sec:pulse_spec}, and the thin dashed line is the
   extrapolation of this fit, below 2 keV.  Please note that the
   data have been rebinned to contain fewer bins than the data
   presented in Table~\ref{tab:pnspectralfits}, to make the plots
   clearer.  The best fit is, however, the same as that presented in
   Table~\ref{tab:pnspectralfits}.}  \label{fig:pulsarspec}
\end{figure}

The PN spectra were extracted using regions as described in
Sect.~\ref{sec:pntiming}.  We used the PN `timing and burst
redistribution matrix file' supplied by the XMM-Newton Science
Operations
Centre\footnote{http://xmm.vilspa.esa.es/external/xmm\_sw\_cal/calib\_frame.shtml}.
Again we used Xspec to fit the spectra and we binned the data to have
at least 50 counts/bin.  The PN spectrum, between 0.6-10.0 keV, is
well fitted by a single absorbed power law of photon index
1.08$\pm$0.06 (see Fig.~\ref{fig:pulsarspec}, left hand side), in good
agreement with the MOS data (1.10$\pm$0.06).  The photon index of the
power law fit is the same, within the errors, as that found by
\cite{mine00} (0.94$\pm$0.22), who fitted the spectrum of \object{PSR
J0218+4232}, observed by BeppoSAX, between 2-10 keV.  

\begin{table*}
\begin{minipage}{18cm}
\caption{Best fitting models to  spectra from the MOS and PN data.  In all the fits where no instrument is mentioned, the data is from the PN.  Regions fitted are the whole lightcurve, the whole of the pulsed region, each of the pulses individually, the interpulse and the non-pulsed region (N$_{\hbox{\rm H}}$=5 $\times 10^{20} {\rm cm}^{-2}$).  The upper part of the table shows the data minus the background.  The lower part of the table shows the data minus the non-pulsed region and the background.  The flux given is unabsorbed flux (${\rm ergs\ cm}^{-2} {\rm s}^{-1}$) in the 0.6-10.0 keV range, averaged over the whole of the pulse period, with errors of the order $\pm$10\%.}
\label{tab:pnspectralfits}
\begin{center}
\begin{tabular}{lccccccccc}
\hline
Region & Phase & Model & kT$_1$ (keV) & kT$_2$ (keV) & Photon Index &
$\chi^{\scriptscriptstyle 2}_{\scriptscriptstyle \nu}$ & dof & Flux
($\times 10^{-13} $)\\
\hline
{\bf All (MOS 1 \& 2)} & 0.00-1.00 & PL & & & 1.10$\pm$0.06 &  1.00 & 29 & 3.9 \\
 & & BB + PL & 0.29$\pm$0.09 & & 0.82$\pm$0.22 &  0.96 & 27 & \\
{\bf All (PN)} & 0.00-1.00 & PL & & & 1.08$\pm$0.06 &  0.93 & 39 & 3.6 \\
 & & BB + PL & 0.19$\pm$0.12 & & 1.08$\pm$0.14 &  0.99 & 37 & \\
{\bf Pulsed} & 0.58-1.45 & PL & & & 1.19$\pm$0.06 &  0.99 & 41 & 3.2\\
 & & 2 BB & 0.23$\pm$0.19 & 1.12$\pm$0.15 &  & 1.03 & 39 & \\
{\bf Pulse 1} & 0.11-0.45 & PL & & & 1.20$\pm$0.07 &  0.93 & 32 & 1.7 \\
 & & BB + PL & 0.29$\pm$0.06 & & 0.80$\pm$0.24 &  0.85 & 30 & \\
 & & 2 BB & 0.33$\pm$0.03 & 1.83$\pm$0.24 &  & 0.94 & 30 & \\
{\bf Pulse 2} & 0.58-0.88 & PL & & & 1.15$\pm$0.09 &  0.74 & 22 & 1.1\\
 & & 2 BB & 0.19$\pm$0.05 & 1.04$\pm$0.08 &  & 0.59 & 20 & \\
{\bf Interpulse} & 0.88-1.11 & PL & & & 1.73$\pm$0.33 &  0.79 & 6 & 0.4\\
 & & BB + PL & 0.08$\pm$0.05 & & 1.34$\pm$0.49 &  0.56 & 4 & \\
{\bf Non-pulsed} & 0.45-0.58 & PL & & & 1.17$\pm$0.37 &  0.55 & 5 & 0.2\\
 & & BB & 0.68$\pm$0.14 & &  & 0.77 & 5 & \\
\hline
{\bf Pulsed} & 0.58-1.45 & PL & & & 1.14$\pm$0.13 &  0.99 & 41 & 2.7\\
 & & BB + PL & 0.22$\pm$0.10 & & 0.82$\pm$0.40 &  0.99 & 39 & \\
 & & 2 BB & 0.25$\pm$0.05 & 1.57$\pm$0.31 &  & 0.89 & 39 & \\
{\bf Pulse 1} & 0.11-0.45 & PL & & & 1.15$\pm$0.13 &  1.13 & 32 & 1.5\\
 & & BB + PL & 0.26$\pm$0.07 & & 0.55$\pm$0.43 &  1.07 & 30 & \\
 & & 2 BB & 0.29$\pm$0.04 & 1.98$\pm$0.53&  & 1.04 & 30 & \\
{\bf Pulse 2} & 0.58-0.88 & PL & & & 1.05$\pm$0.20 &  0.87 & 22 & 0.8\\
\hline

\end{tabular}

PL = power law, BB = blackbody
\end{center}
\end{minipage}
\end{table*}

The advantage of the PN data is that we can fit the spectra from
different regions in the lightcurve.  In all the following fits, we
assume N$_{\hbox{\rm H}}$ = 5 $\times 10^{20} {\rm cm}^{-2}$ and we simply
subtracted a fraction of the background spectrum (as above)
corresponding to the fraction of the pulsar phase interval that we
were interested in, unless stated otherwise.  

\subsubsection{The pulsed region}
\label{sec:pulse_spec}

We first extracted the spectrum of the whole of the pulsed region
($0.58 \leq \phi < 1.45$, see Fig.~\ref{fig:folded}) and binned it up
to have at least 30 counts/bin.  The results of the best spectral fits
and the unabsorbed flux of this pulse can be found in
Table~\ref{tab:pnspectralfits}.  We then subtracted the non-pulsed
region from the pulsed region data, as
\cite{mine00}, so that we could study the pulsed emission only.  We
find, that a power law gives a good fit to the data.  The spectrum
with the best power law fit, can be seen in Fig.~\ref{fig:pulsarspec},
right hand side.  However, we find that the power law photon index of
the pulsed emission (1.19$\pm$0.06) is softer than that found by
\cite{mine00} (0.61$\pm$0.32), who fitted the BeppoSAX data between
$\sim$2-10 keV, even taking into account the errors.  The main
difference between the XMM-Newton data and the BeppoSAX data is that
XMM-Newton has spectral information down to $\sim$0.6 keV, whereas the
BeppoSAX spectral information has a lower limit of about 2 keV.  This
discrepancy in the photon indices of the power law fits to the data
sets may therefore stem from an excess of low energy photons in the
XMM-Newton data, not observable by BeppoSAX.  We therefore tried to
fit our own data in the same energy band as
\cite{mine00}. Using only this energy band we found that a power law,
with a photon index of 0.68$\pm$0.32 ($\chi^{\scriptscriptstyle
2}_{\scriptscriptstyle \nu}$=0.99, 15 dof), gave a good fit to the
data.  This power law fit can be seen as the thick dashed line in
Fig.~\ref{fig:pulsarspec}, right hand side.  The photon index is
consistent with the power law photon index found by \cite{mine00}.
The thin dashed line indicates the extrapolation of this power law fit
to lower energies.  However, the data points at low energies (0.6-2
keV) deviate from the extrapolation of the power law fit.  Including a
blackbody model, where such thermal emission may be expected following
theoretical work e.g. \cite{zhan03}, gives an equally good fit to the
data as a single power law fit.  We also find that the power law
photon index is now compatible with the photon index found by
\cite{mine00}, with the BeppoSAX data.  A two component blackbody
model also gives a good fit to the data, see
Table~\ref{tab:pnspectralfits}.

If we assume that the spectrum above 2 keV is best fitted by the power
law given in
\cite{mine00} (as we have ruled out the single blackbody fit and also
shown that we find a similar fit for our data above 2 keV), we can use
this value to try to determine whether the single power law or a
blackbody plus a power law gives the better fit to the data.  Fitting
the pulsed spectrum with the power law fixed at the \cite{mine00}
value, we find $\chi^{\scriptscriptstyle 2}_{\scriptscriptstyle
\nu}$=1.79 (43 dof) and doing the same for the two component model, we
find $\chi^{\scriptscriptstyle 2}_{\scriptscriptstyle \nu}$=1.17 (41
dof) (blackbody, kT=0.18$\pm$0.04).  Using an F-test we found that the
improvement in the fit was significant at the 3.8$\sigma$ level.  Even
if we used the softest power law slope possible, taking into account
the error on the \cite{mine00} data, we found that the improvement in
the fit was significant at the 2.6$\sigma$ level.  Thus we may have
some indication that the two component model may be the best fit to
the data.

%A comparison of the pulsar spectrum and the pulsed spectrum (minus the
%non-pulsed data) can be seen in Fig.~\ref{fig:pulsarspec}.

\subsubsection{Pulse 1}
\label{sec:pulse1_spec}

\begin{figure}
   \centering \includegraphics[angle=0,width=9cm]{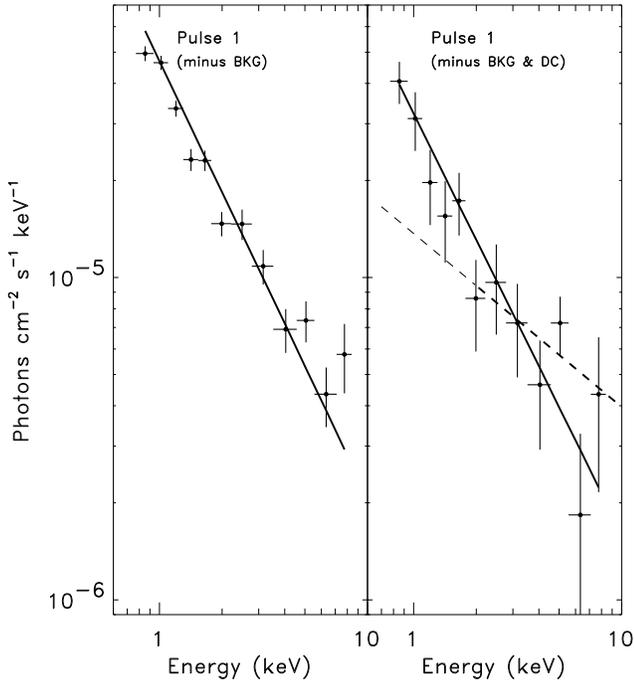}
   \caption{Plots showing the power law fit (solid
   line) to pulse 1.  On the
   left is pulse 1 minus the background contribution and on the right
   is pulse 1 minus the DC and background contributions.  The thick
   dashed line is the best fit to the data above 2 keV, see
   Section~\ref{sec:pulse1_spec}, and the thin dashed line is the
   extrapolation of this fit to below 2 keV.  These data are not
   averaged over the whole of the period, so the exposure time =
   $\Delta\phi$.~T (where $\Delta\phi$ is the phase over which the
   pulse was extracted).  Please note that the data have been rebinned
   to contain fewer bins than the data presented in
   Table~\ref{tab:pnspectralfits}, to make the plots clearer.  The
   best fit is, however, the same as that presented in
   Table~\ref{tab:pnspectralfits}.}  \label{fig:pulsebbodypl}
\end{figure}

We then extracted the spectrum from pulse 1 only ($0.11 \leq \phi <
0.45$), binned it up to have at least 30 counts/bin and subtracted the
background.  The results of the best spectral fits and the unabsorbed
flux of this pulse can be found in Table~\ref{tab:pnspectralfits}.
Using the F-test we find that all the models that we have tried are
equally probable as the most appropriate fit.  The data from pulse 1
plotted with the best power law fit can be found in
Fig.~\ref{fig:pulsebbodypl} (left hand side).  A single blackbody and
a thermal bremsstrahlung model give poor fits to the data.
Subtracting the non-pulsed region from the pulse 1 data, as before, we
find similar spectral fits to those found with the simple background
extraction.  The data plotted with the best power law fit can be seen
in Fig.~\ref{fig:pulsebbodypl}, right hand side.  The power law
spectral index that we find using the XMM-Newton data between 0.6-10.0
keV (1.15$\pm$0.13) is larger than that found by \cite{mine00} (2-10
keV), who found 0.84$\pm$0.35, although it is compatible within the
error bars.  We did, however, try to fit our data above 2 keV, as in
Section~\ref{sec:pulse_spec}.  We found that a harder power law with
photon index 0.66$\pm$0.29 gave a good fit to the data,
$\chi^{\scriptscriptstyle 2}_{\scriptscriptstyle \nu}$=1.29 (19 dof).
This is again compatible with the \cite{mine00} result.  The power law
fit above 2 keV can be seen as the thick dashed line in
Fig.~\ref{fig:pulsebbodypl}, right hand side.  The thin dashed line
indicates the extrapolation of this power law fit to lower energies.
 
\subsubsection{Pulse 2}

\begin{figure}
   \centering
   \includegraphics[angle=0,width=9cm]{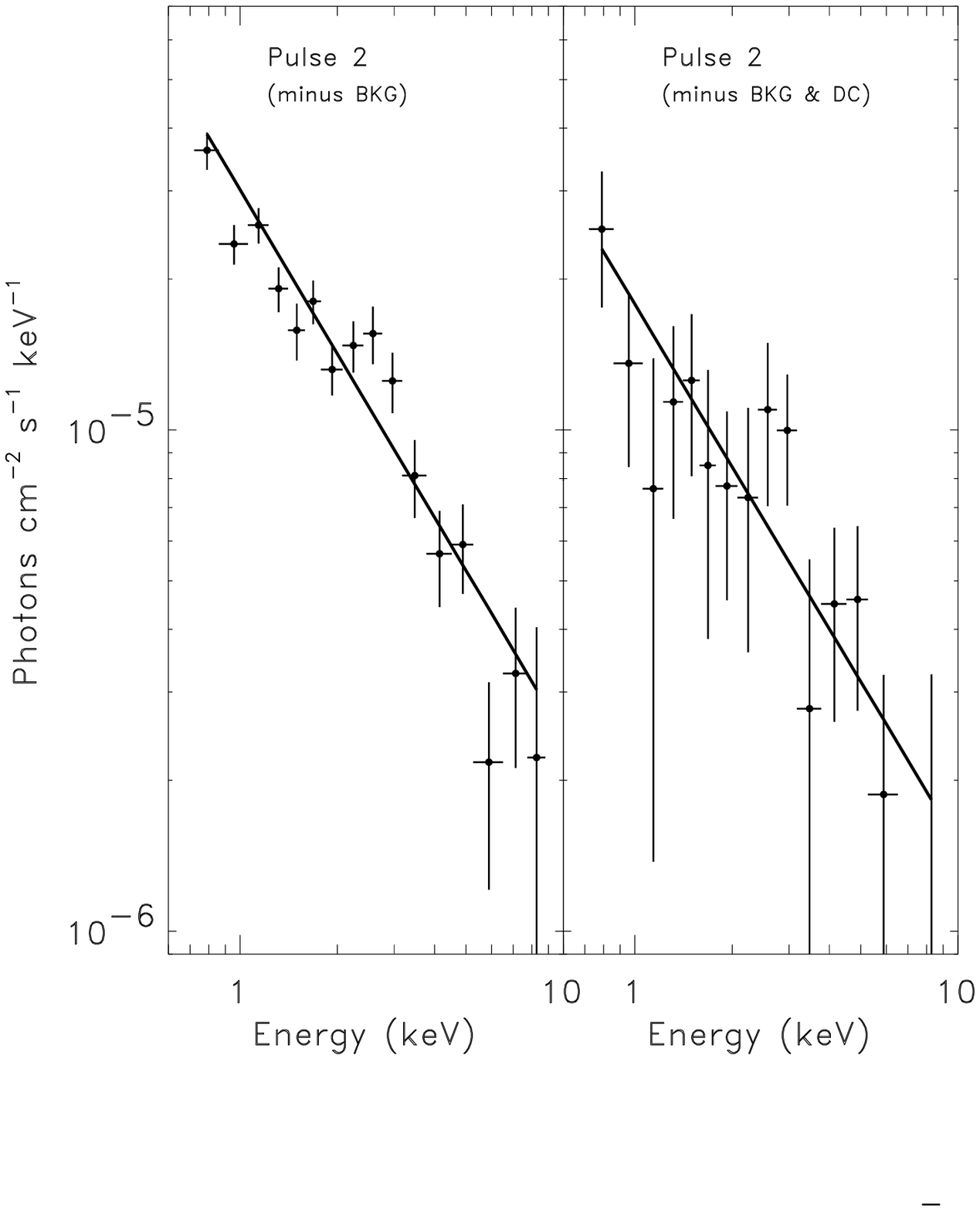}
      \caption{Plots showing the best fit (power law, solid line) to pulse 2.  On the left is pulse 2 minus the  background contribution and on the right is pulse 2 minus the DC and background contributions.  These data are not averaged over the whole of the period, so the exposure time = $\Delta\phi$.~T (where $\Delta\phi$ is the phase over which the pulse was extracted).  Please note that the data have been rebinned to contain fewer bins than the data presented in Table~\ref{tab:pnspectralfits}, to make the plots clearer.  The best fit is, however, the same as that presented in Table~\ref{tab:pnspectralfits}.}
   \label{fig:smallpulsebbodypl}
\end{figure}

Extracting the spectrum from pulse 2 only ($0.58 \leq \phi < 0.88$),
binning it up to have at least 20 counts/bin and subtracting the
background, we find the fits presented in
Table~\ref{tab:pnspectralfits}.  Again a single blackbody and a
thermal bremsstrahlung model give poor fits to the data. The data
plotted with the power law model can be found in
Fig.~\ref{fig:smallpulsebbodypl} (left hand side).  We subtracted the
non-pulsed data from the pulse 2 data (as above). We again find
similar results for the power law fit, as with the simple background
extraction, however, as we found with the pulsed emission, the power
law photon index is softer than that found by \cite{mine00}
(0.42$\pm$0.36), again taking into account the errors.  These data and
the power law model can be found in Fig.~\ref{fig:smallpulsebbodypl}
(right hand side).

\subsubsection{The interpulse region}

Fitting the spectrum to the interpulse region ($0.88 \leq \phi <
1.11$) in a similar way, we find the fits as given in
Table~\ref{tab:pnspectralfits}.  The errors are much larger in this
case, due to the small number of bins.  Again a single blackbody and a
thermal bremsstrahlung model give poor fits to the data.  We
subtracted the non-pulsed data from the interpulse data (as above),
however this resulted in too few counts to create a spectrum.

\subsubsection{The non-pulsed region}

Fitting
the spectrum of the non-pulsed region ($0.45 \leq \phi < 0.58$) in a
similar way (results in Table~\ref{tab:pnspectralfits}), we find that
a simple power law gives the best fit to the data.  In this case, a
blackbody plus a power law model does not give a good fit.  Taking the
non-pulsed flux (0.6-10.0 keV) given in
Table~\ref{tab:pnspectralfits}, and the total flux (0.6-10.0 keV) also
given in Table~\ref{tab:pnspectralfits}, we find a pulsed fraction of
61$^{\scriptscriptstyle +7}_{\scriptscriptstyle -5}$\%, consistent
with the pulsed fraction given in Table~\ref{tab:pulsedfracs}
(0.6-12.0 keV).

\section{Discussion}

Thanks to the large collecting area of XMM-Newton \citep{jans01}, even
from a short observation of \object{PSR J0218+4232}, we are able to
undertake detailed spectral analysis of the various components of the
lightcurve.  

\object{PSR J0218+4232} shows a very hard spectrum and an analogy is
often made between this MSP and two others, \object{PSR B1821-24} and
\object{PSR B1937+21}, which also show pulsed emission, large \.{E}
values (2-20 $\times$ 10$^{35}$ ergs s$^{-1}$) and X-ray spectra with
photon indices less than 2 \citep[][]{sait97,taka01}.  These pulsars
are in turn compared to the Crab pulsar, which shows narrow pulsations
and a hard X-ray spectrum \citep{toor77}.  It is believed that such
hard emission arises from the magnetosphere.  However, the Crab pulsar
shows pulses which have FWHM of $\sim$0.042 and 0.075 in phase at
similar X-ray energies \citep{eike97}, which are 2-3 times narrower
than those of \object{PSR J0218+4232}.  \object{PSR B1821-24} also has
a pulse width of less than 0.04 in phase \citep[][]{rots98,sait97} and
those of \object{PSR B1937+21} are also only $\sim$0.06 in phase
\citep[][]{taka01}.  The larger width of the \object{PSR J0218+4232}
pulses may simply be due to it being an aligned rotator
\citep{nava95}.  However, \cite{luo00} state that with their polar cap
model, they can reproduce the wide double-peaked profile without
assuming an aligned rotator.  Nonetheless, the particularly hard
spectrum, extending to $\gamma$-rays \citep{kuip02b,kuip00} and
the wider pulses shown by \object{PSR J0218+4232} could indicate that
it is a different kind of MSP to \object{PSR B1821-24} and \object{PSR
  B1937+21}.

\subsection{The pulsed emission}
\label{sec:PNpulse1}

We have seen that the form of the pulsed emission varies significantly
with energy (see Sect.~\ref{sec:pntiming} and
Fig.~\ref{fig:psrdiffenergy}).  We have therefore tried a variety of
simple models to fit the pulsed emission data, see
Sect.~\ref{sec:specan} and Table~\ref{tab:pnspectralfits}.  For the
pulsed spectrum and the spectra of pulses 1 and 2, we find that a
single blackbody fit is not a good description of the data, but that a
single power law does give a good fit.  However, the photon indices
that we find when fitting the XMM-Newton data, are significantly
softer than those found by \cite{mine00}, when fitting the spectra
observed with the BeppoSAX MECS between 2 and 10 keV.  By fitting the
XMM-Newton spectra with a two component model (blackbody plus power
law, in the same way as for PSR J0437-4715 \citep{zavl02}) we recover
the same power law photon indices found by \cite{mine00}, which could
indicate that there may be additional emission below 2 keV and
therefore not observable with BeppoSAX.  Taking the blackbody
temperature (2.90$\pm$0.70 $\times 10^{6}$ K) from the blackbody plus
power law fit to pulse 1, we find a radius of emission of
0.37$\pm$0.33 km (90\% confidence limit), consistent with the
temperature and radius of a polar cap
\citep[10$^6$-10$^7$ K and $\sim$1 km e.g.][and references
therein]{zhan03,zavl98}.  

There is some evidence from both the spectral fitting and the
lightcurves presented in Sect.~\ref{sec:pntiming} that the interpulse
region emits a softer spectrum than that of the two pulses \citep[as
also seen by][]{kuip02}.

%Pulse 2 shows little evidence for a blackbody component, where it
%appears to be best fitted by a simple power law.

%The blackbody temperatures and power law photon indices are
%similar, within the errors, for each of these regions.  The photon
%indices determined with the blackbody plus power law model are
%similar, within the errors, to those determined by \cite{mine00}, who
%fitted the spectra observed by BeppoSAX in the 2.0-10.0 keV region
%only.  

%This could indicate that The
%non-thermal radiation thus originates in the pulsar magnetosphere,
%which is supported by the fact that \object{PSR J0218+4232} is also
%detected at high energies.

%Fitting pulse 2 with a two component blackbody indicates that the two
%temperatures are slightly lower and therefore the radii are slightly
%larger.

\subsection{The non-pulsed emission}
\label{sec:nonpulsed}

With regards to the non-pulsed region, we find that a simple
blackbody, with a temperature consistent with that determined using
the ROSAT HRI, BeppoSAX MECS and Chandra instruments \citep{kuip02},
can fit the data.  However, the radius of emission determined from
this blackbody is only 0.10$\pm$0.08 km.  This indicates that
from the small radius, the emission is unlikely to be from a polar
cap. \cite{kuip02} suggest that this could nonetheless be thermal
emission from the polar caps, if \object{PSR J0218+4232} is a nearly
aligned rotator and the viewing angle is small, so that the polar caps
are always in sight.  Alternatively, the data can be well described by
a single power law, where the photon index is the same, within the
errors, as is found to fit all the other sections of the lightcurve.
This would be consistent with recent polar cap models, such as that of
\cite{zhan03}, where X-ray emission in these models consists of one
power law and two thermal components.  In these cases the nonthermal
X-rays are produced by the synchrotron radiation of e$^{\pm}$ pairs
created in a strong magnetic field near the stellar surface; the soft
thermal X-rays are produced by heating of polar cap areas with a
radius of the order of 1 km; and the medium hard thermal X-rays result
from the polar cap heating by the return particles from the outer gap.
This is consistent with data taken at higher energies
\citep[see][]{kuip00,kuip02b}, where \cite{kuip00} determine that
extrapolating a hard power law spectrum from the XMM-Newton spectral
range is just in agreement with the OSSE upper limit.

\section{Conclusions}

We have presented XMM-Newton MOS imaging (0.2-10.0 keV) and PN timing
(0.4-12.0 keV) data of \object{PSR J0218+4232}.  Using the large
number of counts, we have confirmed the previously detected pulsations
of \object{PSR J0218+4232} and we have shown that the form of the
lightcurve varies with energy.  We have shown that the broad band
X-ray spectrum between 0.6-10.0 keV of the pulsed emission is softer
than the spectrum measured between 2-10 keV by BeppoSAX, and we have
shown some evidence that a simple power law may not be a good
description of the 0.6-10.0 keV spectrum.  A longer observation of
this millisecond pulsar should establish whether or not this is indeed
the case.

\begin{acknowledgements}
  
  We would like to thank G. Ramsay, U. Lammers and M. Kirsch for their
  help in analysing the timing anomalies of the PN data and B. Gendre
  and P. Jean for their technical advice.  We are also grateful to
  W. Klu\'zniak and J.  Poutanen for several enlightening discussions.

\end{acknowledgements}


\begin{thebibliography}{}
  
%\bibitem[Becker \& Tr\"umper (1993)]{beck93} Becker~W., Tr\"{u}mper~J.,
%  1993, Nature, 365, 528
  
%\bibitem[Becker \& Aschenbach (2002)]{beck02} Becker~W.,
%  Aschenbach~B., 2002, Proceedings of the 270. WE-Heraeus Seminar on
%  Neutron Stars, Pulsars, and Supernova Remnants. MPE Report 278. Ed.
%  W. Becker, H. Lesch, J. Tr\"umper.

\bibitem[Cheng et al. (2000)]{chen00} Chen~K.~S., Ruderman~M.,  Zhang~L., 
  2000, ApJ, 537, 964


\bibitem[Dwarakanath \& Shankar (1990)]{dwar90} Dwarakanath~K.S.,
  Shankar~N.U., 1990, A\&A, 11, 323
  
\bibitem[Eikenberry \& Fazio (1997)]{eike97} Eikenberry~S.S.,
  Fazio~G.G., 1997, ApJ, 476, 281
  
\bibitem[Hales et al. (1993)]{hale93} Hales~S.~E.~G., Baldwin~J.E.,
  Warner~P.J., 1993, MNRAS, 263, 25
  
\bibitem[Harding \& Muslimov (2002)]{hard02} Harding~A.~K., Muslimov~A.G., 
  2002, Apj, 568, 862
  
\bibitem[Hasinger et al. (2001)]{hasi01} Hasinger~G., Altieri~B.,
  Arnaud~M., et al., 2001, A\&A, 365, L45
  
\bibitem[Jansen et al. (2001)]{jans01} Jansen~F., Lumb~D., Altieri~B.,
  et al., 2001, A\&A, 365, L1-6
  
\bibitem[Kirsch et al. (2002)]{kirs02} Kirsch~M., and the EPIC
  Consortium, 2002, XMM-SOC-CAL-TN-0018

\bibitem[Kirsch et al. (2003)]{kirs03} Kirsch~M.G.F., Becker~W., Benlloch-Garcia~S., et al., 2003, Proc. SPIE, 5165 

\bibitem[Kuiper et al. (1998)]{kuip98} Kuiper~L., Hermsen~W., Verbunt~F.,
  Belloni~T., 1998, A\&A, 336, 545
  
\bibitem[Kuiper et al. (2000)]{kuip00} Kuiper~L., Hermsen~W., Verbunt~F.,
  Thompson~D.J., Stairs~I.H., Lyne~A.G., Strickman~M.S., Cusumano~G.,
  2000, A\&A, 359, 615
  
\bibitem[Kuiper et al. (2002a)]{kuip02} Kuiper~L., Hermsen~W., Verbunt~F.,
  Ord~S., Stairs~I.H., Lyne~A.G., 2002a, ApJ, 577, 917
  
\bibitem[Kuiper et al. (2002b)]{kuip02b} Kuiper~L., Hermsen~W., 2002b,
  "The Gamma-Ray Universe", Eds. A.~Goldwurm, D.~Neumann,
  J.~Tran~Thanh~Van, The Gioi Publishers (Vietnam)
  
  
\bibitem[Mineo et al. (2000)]{mine00} Mineo~T., Cusumano~G., Kuiper~L.,
  Hermsen~W., Massaro~E., Becker~W., Nicastro~L., Sacco~B.,
  Verbunt~F., Lyne~A.G., Stairs~I.H., Shibata~S., 2000, A\&A, 355,
  1053
  
\bibitem[Luo et al. (2000)]{luo00} Luo~Q., Shibata~S., Melrose~D.B.,
  2000, MNRAS, 318, 943
  
\bibitem[Navarro et al. (1995)]{nava95} Navarro~J., de Bruyn~A.G.,
  Frail~D.A., Kulkarni~S.R., Lyne~A.G., 1995, ApJ, 455, L55
  
\bibitem[Pavlov et al. (1992)]{pavl92} Pavlov~G.G., Shibanov~Y.A.,
  Zavlin~V.E., 1992, MNRAS, 253, 193

\bibitem[Pavlov \& Zavlin (1997)]{pavl97} Pavlov~G.G., Zavlin~V.E., 
  1997, ApJ, 490, 91
  
\bibitem[Rots et al. (1998)]{rots98} Rots~A.H., Jahoda~K., Macomb~D.J.,
  Kawai~N., Saito~Y., Kaspi~V.M., Lyne~A.G., Manchester~R.N.,
  Backer~D.C., Somer~A.L., Marsden~D., Rothschild~R.E., 1998, ApJ,
  501, 749

\bibitem[Rudak \& Dyks (1999)]{ruda99} Rudak~B.,  Dyks~J., 1999, MNRAS, 
  303, 477 

\bibitem[Saito et al. (1997)]{sait97} Saito~Y., Kawai~N., Kamae~T.,
  Shibata~S., Dotani~T., Kulkarni~S.R., 1997, ApJ, 477, 37

\bibitem[Stairs et al. (1999)]{stai99} Stairs~I.H., Thorsett~S.E.,
  Camilo~F., 1999, ApJSS, 123, 627
  
\bibitem[Str\"{u}der et al. (2001)]{stru01} Str\"{u}der~L., Briel~U.,
  Dennerl~K., et al., 2001, A\&A, 365, L18
  
\bibitem[Swanepoel et al. (1996)]{swan96} Swanepoel~J.W.H., De Beer~C.F., 
  Loots~H., 1996, ApJ, 467, 261
  
\bibitem[Takahashi et al. (2001)]{taka01}Takahashi~M., Shibata~S.,
  Torii~K., Saito~Y., Kawai~N., Hirayama~M., Dotani~T., Gunji~S.,
  Sakurai~H., Stairs~I.H., Manchester~R.N., 2001, ApJ, 554, 316
  
\bibitem[Tennant et al. (2001)]{tenn01} Tennant~A.F., Becker~W.,
  Juda~M., Elsner~R.F., Kolodziejczak~J.J., Murray~S.S., O'Dell~S.L.,
  Paerels~F., Swartz~D.A., Shibazaki~N., Weisskopf~M., 2001, ApJ,
  554, L173
  
\bibitem[Toor \& Seward (1977)]{toor77} Toor~A., Seward~F.D., 1977,
  ApJ, 216, 560


\bibitem[Turner et al. (2001)]{turn01} Turner~M.J.L., Abbey~A., Arnaud~M.
  et al., 2001, A\&A, 365, L27
  
\bibitem[Verbunt et al. (1996)]{verb96} Verbunt~F., Kuiper~L.,
  Belloni~T., Johnston~H.M., de Bruyn~A.G., Hermsen~W., van der
  Klis~M., 1996, A\&A, 311, L9
  
\bibitem[Zavlin et al. (1996)]{zavl96} Zavlin~V.E., Pavlov~G.G., Shibanov~Y.A.,
  1996, A\&A, 315, 141

\bibitem[Zavlin \& Pavlov (1998)]{zavl98} Zavlin~V.E., Pavlov~G.G.,
  1998, A\&A, 329, 583

\bibitem[Zavlin et al. (2002)]{zavl02} Zavlin~V.E., Pavlov~G.G.,
  Sanwal~D., Manchester~R.N., Tr\"{u}mper~J., Halpern~P., Becker~W.,
  2002, ApJ, 569, 894
  
\bibitem[Zhang \& Cheng (2003)]{zhan03} Zhang~L., Cheng~K.S., 2003,
  A\&A, 398, 639

\end{thebibliography}
\end{document}